\documentclass[aps,pra,reprint,nofootinbib,showpacs,superscriptaddress]{revtex4-1}
\usepackage{amsmath}
\usepackage{amsfonts}
\usepackage{amssymb}
\usepackage{amstext,bm}
\usepackage{textcomp}
\usepackage[colorlinks=true,linkcolor=blue,citecolor=red,plainpages=false,pdfpagelabels]%
{hyperref}
\usepackage{color}
\usepackage[makeroom]{cancel}
\usepackage{hyperref}
\usepackage{braket}
 \def\be{\begin{equation}}
 \def\ee{\end{equation}}
 \def\bes{\begin{eqnarray}}
 \def\ees{\end{eqnarray}}
\begin{document}

\title
{Continuous-variable quantum Gaussian process regression and quantum singular value decomposition of non-sparse low rank matrices}
\author{Siddhartha Das}
\affiliation{Hearne Institute for Theoretical Physics,
	Louisiana State University,
	Baton Rouge, LA 70803, USA}
\author{George Siopsis}
\affiliation{Department of Physics and Astronomy,
	The University of Tennessee, Knoxville, TN 37966-1200, USA}
	\author{Christian Weedbrook}
	\affiliation{Xanadu, 372 Richmond St W, Toronto, M5V 2L7, Canada}
\date{\today}

\begin{abstract}
With the significant advancement in quantum computation in the past couple of decades, the exploration of machine-learning subroutines using quantum strategies has become increasingly popular. Gaussian process regression is a widely used technique in supervised classical machine learning. Here we introduce an algorithm for Gaussian process regression using continuous-variable quantum systems that can be realized with technology based on photonic quantum computers under certain assumptions regarding distribution of data and availability of efficient quantum access. Our algorithm shows that by using a continuous-variable quantum computer a dramatic speed-up in computing Gaussian process regression can be achieved, i.e., the possibility of exponentially reducing the time to compute. Furthermore, our results also include a continuous-variable quantum-assisted singular value decomposition method of non-sparse low rank matrices and forms an important subroutine in our Gaussian process regression algorithm. 
\end{abstract}
\maketitle

\section{Introduction}
One of the current technological needs in the area of computer science is finding an efficient and faster way of manipulating large data sets, and extracting worthwhile inferences. In the last decade, machine-learning techniques have been used to perform many tasks involving big data. In 1959, Arthur Samuel defined machine learning as the \textquotedblleft field of study that gives computers the ability to learn without being explicitly programmed\textquotedblright \cite{Sam59}. Machine learning has not only helped us better understand the human genome, but has also made self-driving cars, practical speech recognition, effective web search, etc., possible \cite{LN15,LAB+11}.

One of the two machine-learning methods involves supervised learning (with the other task being unsupervised learning)~\cite{RW05book}. It is the problem of learning input-output mappings from an empirical (training) data set. Depending on the nature of the output, the problem of supervised learning can be categorized under two types: regression and classification. Regression deals with the process involving continuous output, whereas classification deals with the process involving discrete (categorical) output. 

Under supervised learning, one is given a data set $\mathcal{D}$ containing $n$ observations of input-output $(\bm{x}_i,y_i)$, where $ \bm{x}_i\in\mathbb{R}^d,y_i\in\mathbb{R}$, $\forall\ i\in\{0,1,\ldots,N-1\}$. This is a training data set involving a process called regression as one deals with continuous output. Given this training data set, the machine is trained to predict new inputs which are not listed in $\mathcal{D}$. The goal of supervised learning is to induce a function from observations on the training data set.

Gaussian processes form powerful models for regression problems. They have found a wide range of applications: robotics, data mining, geophysics, climate modeling, etc.\ (see \cite{RW05book} and references therein). Any Gaussian distribution is fully characterized by its mean and covariance function. The problem of learning in a Gaussian process is precisely the problem of finding suitable properties of the covariance function. In general, when only classical systems and strategies are in use, the implementation of a Gaussian process regression model with $n$ training points typically requires $\mathcal{O}(n^3)$ basic operations~\cite{RW05book}. 

The application of principles in quantum mechanics has led to the realization of technologies in information processing and computation that can never be achieved within the realm of classical mechanics, such as teleportation and quantum key distribution~\cite{NC00, BD00, Dow13book}. Significantly, quantum computers are expected to have advantages over classical computers~\cite{NC00}. In theoretical computer science, quantum algorithms have been developed showing a significant advantage over their classical counterparts in terms of, in the best case scenario, an exponential speedup~\cite{Mon15}. Importantly, quantum algorithms can also have a significant impact on machine learning, and this has led to the emergence of quantum machine learning~\cite{BWP+16}. 

The HHL algorithm, introduced in~\cite{HHL09}, gives a quantum algorithm for solving systems of linear equations. Specifically, let the system of linear equations be $A\bm{x}=\bm{b}$, where $A$ is a matrix, and $\bm{b}$ a vector, and the goal is to find the vector $\bm{x}$. In~\cite{HHL09}, the case was considered in which one needs to know the expectation value of some operator associated with $\bm{x}$, e.g., $\bm{x}^\dag M\bm{x}$ for a given matrix $M$, instead of the solution $\bm{x}$ itself. Assuming $A$ is a sparse $N\times N$ matrix with condition number (ratio of largest and smallest eigenvalue) $\kappa$, classical algorithms can find $\bm{x}$ and estimate $\bm{x}^\dag M\bm{x}$ in $\mathcal{O}(N\sqrt{\kappa})$ time. However, in~\cite{HHL09} the authors presented a quantum algorithm that ran in $\textnormal{poly}(\log N,\kappa,1/\epsilon)$ time, with $\epsilon$ precision in the output state (N.B., if one, in certain cases, avoids phase estimation the precision can be $\textnormal{poly}(\log (1/\epsilon))$~\cite{Childs2015}).
They showed that when the sparseness parameter of the matrix does not scale faster than polylogarithmically in $N$, an exponential speedup is possible with the quantum linear systems algorithm. Recently, this quantum algorithm was applied to Gaussian process regression~\cite{ZFF15}. One of the contexts in which sparsely constructed Gaussian processes find applications is the problem involving inference in large data sets~\cite{MR09}. However, all applications so far have been limited to qubit or discrete-variable (DV) quantum systems.  

Continuous-variable (CV) quantum systems are characterized by having an infinite-dimensional Hilbert space, and measurements involving observables with continuous eigenspectra \cite{WPG+12}. A CV generalization of any DV quantum system assisted algorithm is essential in the context of developing algorithms for quantum computers involving CV systems, e.g., optical quantum computing \cite{WPG+12}. The usefulness of CV quantum machine learning~\cite{LPSW17} goes beyond the processing of classical data sets that involve a discrete number of data. The output by the universal CV quantum computation is a CV state that evolves under a designed Hamiltonian~\cite{LB99}. The DV machine learning subroutines are inefficient (incapable) of processing full CV states by themselves. This deficit of DV quantum systems assisted machine-learning subroutines can be curbed using predominantly CV quantum systems along with qubits when needed~\cite{LB99}. 

In this paper, we apply the techniques developed in~\cite{LPSW17} to generalize the DV quantum-assisted Gaussian process regression~\cite{ZFF15} to CV systems. For our task, we also describe an encoding method of a covariance matrix that gives a technique for a CV quantum-assisted singular value decomposition method of non-sparse low rank matrices~\cite{RSL16}. Furthermore, we consider the practical case of finite squeezing analysis for our algorithm. 

Our discussion is organized as follows. We first introduce our notation and basic definitions in Section~\ref{sec:pre-GPR}. In Section~\ref{III}, we introduce a CV method of quantum singular value decomposition of non-sparse low rank matrices.  In Section~\ref{IV}, we illustrate a CV quantum system assisted algorithm for a Gaussian process regression model by efficiently computing its mean in \ref{sec:algo-mean} and covariance function in \ref{sec:algo-cov}. We base the algorithm on a scheme to encode the covariance matrix in an oracular setting for an efficient computation of the mean and covariance functions using CV quantum systems. Finally, in Section~\ref{sec:con}, we give concluding remarks.

\section{Classical Gaussian process regression}\label{sec:pre-GPR}

In this review section, we introduce our notation and basic definitions that are needed for the discussion of Gaussian process regression. Let $\mathcal{N} (\bm{x} | \bm{m} , \sigma^2)$ denote the Gaussian (normal) distribution of the variable $\bm{x}$ with mean $\bm{m} = \mathbb{E} [\bm{x}]$ and variance $\mathbb{V}[\bm{x}] = \mathbb{E} [(\bm{x} - \bm{m})^2] = \sigma^2$. Consider a training set $\mathcal{D}=\{\bm{x}_i,y_i\}_{i=0}^{N-1}$ of $N$ $d$-dimensional inputs (input vectors) $\bm{x}_i$ and scalar outputs (or target values) $y_i$ ($i\in\{0,1,\dots N-1\}$). The outputs $y_i$ are accumulated together to form entries of an output $N$-dimensional vector $\bm{y}$. Furthermore, we assume that the outputs are noisy, i.e., 
\be y_i = f(\bm{x}_i) + \varepsilon, \ee
where $f(\bm{x}_i)$ is the latent function~\cite{RW05book} and $\varepsilon \sim \mathcal{N} (0, \sigma^2)$ denotes independent and identically distributed Gaussian noise. 

A Gaussian process is a collection of random variables, any finite number of which have a joint Gaussian distribution. A Gaussian process is completely specified by its mean and covariance function (kernel)~\cite{RW05book}, which for a real
process $f(\bm{x})$ is defined by
\be k(\bm{x}, \bm{x'}) = \mathbb{E} [ (f(\bm{x}) - m(\bm{x})) (f(\bm{x}') - m(\bm{x}'))], \ee
where $m(\bm{x}) = \mathbb{E}[f(\bm{x})]$ denotes the mean function of $f(\bm{x})$. Let us denote the real process as 
\be f(\bm{x})\sim \mathcal{GP}(m(\bm{x}),k(\bm{x},\bm{x}')).\ee
Given a new input (test point) $\bm{x}_\ast$, our goal is to predict the distribution of
\be f_\ast = f(\bm{x}_\ast). \ee One can consider an array of test points, however, for simplicity we consider only a single test point. The procedure described for a single test point can be simply generalized to multiple test point instances.

To this end, we note that the joint distribution of the observed target values and the function value at the test location are, respectively,
\be \left[ \begin{array}{c}
	\bm{y} \\ f_\ast
\end{array} \right] \ \sim \ \mathcal{N} ( \bm{0}, \mathbb{K} ) \ , \ \ \mathbb{K} = \left[ \begin{array}{cc}
K & \bm{k}_\ast \\ \bm{k}_\ast^T & k_{\ast\ast}
\end{array}\right],
\ee
where $K$ is the $N\times N$ matrix with entries \be K_{ij} = k(\bm{x}_i, \bm{x}_j) + \sigma^2 \delta_{ij} \ ,\label{eq:matrix-K} \ee and the entries of the vector $\bm{k}_\ast\equiv [\bm{k}_{\ast i}]$ are the covariance functions $k(\bm{x}_i, \bm{x}_\ast)$, and $k_{\ast\ast} = k(\bm{x}_\ast, \bm{x}_\ast)$. Without loss of generality, we set the mean of the distribution to zero. 

Using
\be \mathbb{K}^{-1} = \left[ \begin{array}{cc}
	K^{-1} + \tilde{k}_{\ast\ast}^{-1} \tilde{\bm{k}}_\ast \tilde{\bm{k}}_\ast^T & -\tilde{k}_{\ast\ast}^{-1}\tilde{\bm{k}}_\ast \\ -\tilde{k}_{\ast\ast}^{-1}\tilde{\bm{k}}_\ast^T & \tilde{k}_{\ast\ast}^{-1}
\end{array}\right], \ee
where
\be \label{eq:variance_formula} \tilde{\bm{k}}_\ast = K^{-1} \bm{k}_\ast \ , \ \  \tilde{k}_{\ast\ast} = k_{\ast\ast} - \bm{k}_\ast\cdot \tilde{\bm{k}}_\ast,
\ee
we deduce the conditional probability to be
\be P(f_\ast | \bm{y} ) \ \sim \ \mathcal{N} ( \bm{y}\cdot \tilde{\bm{k}}_\ast , \tilde{k}_{\ast\ast}).
\ee
The task at hand boils down to the efficient computation of the mean $\bm{y}\cdot \tilde{\bm{k}}_\ast$ and variance $\tilde{k}_{\ast\ast}$.


\section{Quantum singular value decomposition method of non-sparse low rank matrices}\label{III}

In this section, we introduce a CV version of the quantum-assisted singular value decomposition method of non-sparse low rank matrices which was first introduced for qubits in Ref.~\cite{RSL16}. These results will form a subroutine in the next section for the quantum Gaussian process regression algorithm.

We begin by assuming for simplicity that $N= 2^n$ and the matrix $K$ can be encoded as  \be\hat{K} = \left[ \begin{array}{cc}
	K &  \\  & \mathbb{I}
\end{array}\right] \ee in an ensemble of $n+1$ qubits and accessed via oracle calls. 
For the oracle calls, we make use of the method for non-sparse matrices in an oracular setting which requires only one-sparse simulation techniques~\cite{RSL16}. 

To record and access the matrix $\hat{K}$, we first create the one-sparse Hermitian matrix~\cite{RSL16}
\be H = \sum_{x,y=0}^{2N-1} \langle x| \hat{K} | y\rangle |x\rangle\langle y| \otimes |y\rangle\langle x| \ee
whose entries are real numbers. This enlarges the Hilbert space quadratically, but because the matrix $H$ has a single non-vanishing element in each row, its dynamics that can be efficiently approximated. Indeed,
we approximate each non-vanishing element of $H$ by $2\zeta \left[ \frac{\langle x| \hat{K} | y\rangle}{2\zeta}\right]$, where $[l]$ denotes the integer part of $l$. The resulting approximate matrix
\be\label{eq12} \widetilde{H} = 2\sum_{x,y=0}^{2N-1} \left[ \frac{\langle x| \hat{K} | y\rangle}{2\zeta}\right] |x\rangle\langle y| \otimes |y\rangle\langle x| \ee
has entries which are even integers, and $H\approx \zeta \widetilde{H}$, i.e., $\| H - \zeta \widetilde{H} \| \lesssim \zeta$~\cite{BCC+14}. It can be easily decomposed into a sum of matrices each of which has eigenvalues $\pm 1$,
\be\label{eq13} \widetilde{H} = \sum_{j=1}^{j_{\text{max}}} H_j~, \ee
encoding efficiently a good approximation to all the information in the matrix $\hat{K}$.
The sum in \eqref{eq13} contains at most $\mathcal{O} (\| \widetilde{H} \| ) \sim \mathcal{O} (\lambda_{\text{max}} / \zeta)$ terms, where $\lambda_{\text{max}}$ is the largest eigenvalue of $K$. If $\lambda_{\text{max}}$ is independent of the number of qubits, then the complexity of our quantum calculation is independent of $N$. This is not always the case \cite{Aaronson2015}, and depends on data distribution. We will discuss the complexity of the algorithm for a given error and the restrictions on the matrix $K$ imposed by requiring exponential speed up further in Section \ref{IV}.

We form the oracle calls to access $H$,
\be Q = i\sum_{j=1}^{j_{\text{max}}} |j\rangle\langle j| \otimes e^{-i\frac{\pi}{2}H_j}. \ee
Notice that $Q$ is Hermitian as well as unitary, therefore $Q^2 =\mathbb{I}$.

Next we consider how to prepare states containing encodings of $|\bm{y}\rangle$ and $|\bm{k}_\ast\rangle$. We normally encode a $N$-dimensional unit vector $\bm{v}$ by forming the $n$-qubit state $|\bm{v}\rangle = \sum_i v_i |i\rangle$. However, the vectors we are interested in are not unit vectors, and we are also interested in the signs of inner products, not just their absolute values. To encode this additional information, e.g., for $\bm{y}$, we shall use $n+1$ qubits and form the unit vector corresponding to the $2N$-dimensional vector $[y_{i}/ c(y), \sqrt{1- y_{i}^2/c^2(y)}]$, where $c(y) > y_{i}$, for all $i\in\{ 0,1,\ldots,N-1\}$, and encode it on the $(n+1)$--qubit state 
\be\label{eq15} |\bm{y}\rangle = \frac{1}{\sqrt{N}} \sum_{i=0}^{N-1} \left( \frac{y_{i}}{c(y)} |i\rangle + \sqrt{1- \frac{y_{i}^2}{c^2(y)}} |N+i\rangle \right). \ee
One way to prepare such a state is to form a unitary (sequence of rotations) $U_y$ such that $U_y |0\ldots 0 \rangle = |\bm{y}\rangle$. This can be done efficiently, as long as the components of $\bm{y}$ are relatively uniform \cite{Aaronson2015}. In a similar way, we can encode $\bm{k}_\ast$ in the $(n+1)$-qubit state
\be\label{eq16} |\bm{k}_\ast\rangle = \frac{1}{\sqrt{N}} \sum_{i=0}^{N-1} \left( \frac{k_{\ast i}}{c(k_\ast )} |i\rangle + \sqrt{1- \frac{k_{\ast i}^2}{c^2(k_\ast)}} |N+i\rangle \right). \ee

\section{Quantum Gaussian Process Regression}\label{IV}

In this section, we illustrate our quantum algorithm using quantum CV systems to implement the task of efficiently computing the mean $\bm{y}\cdot \tilde{\bm{k}}_\ast$ and variance $\tilde{k}_{\ast\ast}$ for a Gaussian process regression model. We do so by including our previous results of the quantum singular value decomposition as a subroutine.

\subsection{Efficient computation of mean}\label{sec:algo-mean}

Given the $N$-dimensional vectors $\bm{y}$ and $\bm{k}_\ast$ defined above, we form the $4N$-dimensional vector $[y_{i}/ c(y), \sqrt{1- y_{i}^2/c^2(y)},k_{\ast i}/ c(k_\ast), \sqrt{1- k_{\ast i}^2/c^2(k_\ast)}]$, and encode it in the $(n+2)$-qubit state $|\bm{y},\bm{k}_\ast\rangle$, as outlined above. This can be done efficiently with a string of unitary operations, provided the components are relatively uniform \cite{Aaronson2015}, or they have been encoded and stored in a qRAM \cite{GLM08,GLM08b} by a third party.

Evidently, our input state can be written in terms of the states \eqref{eq15} and \eqref{eq16} defined above as
\be\label{eq:yk} |\bm{y},\bm{k}_\ast\rangle \equiv \frac{1}{\sqrt{2}}\left(|\bm{y}\rangle |0\rangle + |\bm{k}_\ast\rangle |1\rangle\right), \ee 
To this state we append two CV resource modes in the 
squeezed state
\be\label{eq18} |\Phi_{R}(\xi)\rangle = \frac{1}{\sqrt{\pi} \xi}\int dq_R d\tilde{q}_R e^{-\frac{1}{2\xi^2} [q_R^2 + \tilde{q}_R^2]} |q_R\rangle |\tilde{q}_R\rangle~, \ee
written in terms of the $q$ quadratures, $q_R$ and $\tilde{q}_R$, respectively, of the resource modes. It is advantageous to make the squeezing parameter $\xi$ as small as technologically feasible,
thus forming the state
\be\label{eq:chi-y-k} |\chi(\bm{y},\bm{k}_\ast,\Phi_{R} (\xi))\rangle \equiv |\bm{y},\bm{k}_\ast\rangle |\Phi_{R} (\xi)\rangle. \ee
Next we apply the unitary
\be\label{equ13} \mathcal{U} = e^{i\gamma \frac{\hat{K}}{4N} \hat{N} p_R \tilde{p}_R}, \ee
where $\hat{N} = \frac{\mathbb{I} - Z}{2}$ is a projection acting on the last qubit of our state \eqref{eq:yk}, $p_R$, $\tilde{p}_R$ are $p$ quadrature operators acting on the resource modes, and $\gamma$ is a parameter that can be adjusted at will.
To implement it, instead of applying an evolution involving $\hat{K}$, we use the quadratically enlarged matrix $H$ (eq.\ \eqref{eq12}) which contains the same information, but leads to simpler dynamics. We will then use the resulting unitary as a generalized exponential swap to apply the desired matrix $\hat{K}$.

To implement the approximation to $H$, $\widetilde{H}$ (eqs.\ \eqref{eq12} and \eqref{eq13}), we need a `fractional' query, $Q^{\delta p_R \tilde{p}_R\hat{N}}$, where $\delta$ is an arbitrary real number. Notice that
\be Q^{\delta p_R \tilde{p}_R\hat{N}} = \frac{1}{2} (\mathbb{I} + Q) \otimes e^{i\frac{\pi\delta}{2} p_R \tilde{p}_R\hat{N}} + \frac{1}{2} (\mathbb{I} - Q) \otimes e^{-i\frac{\pi\delta}{2} p_R \tilde{p}_R\hat{N}}~, \ee
acting on, say, $|\psi\rangle|\phi\rangle$. We append an ancilla qubit in the state $|+\rangle_A$, and then perform a control-$Q$ on $|\psi\rangle$ with the ancilla as control. We obtain
\be \frac{1}{\sqrt{2}} \left( |\psi\rangle|\phi\rangle |0\rangle_A + Q|\psi\rangle|\phi\rangle |1\rangle_A \right). \ee
Then we rotate the ancilla so that $|0\rangle_A \to |+\rangle_A$, $|1\rangle_A \to |-\rangle_A$, which yields
\be \frac{1}{2} \left( (\mathbb{I} + Q)|\psi\rangle|\phi\rangle |0\rangle_A + (\mathbb{I}- Q)|\psi\rangle|\phi\rangle |1\rangle_A \right). \ee
Next we apply the unitary $e^{i\frac{\pi\delta}{2} p_R \tilde{p}_R \hat{N}Z_A}$, where $Z_A$ is the Pauli matrix $Z$ acting on the ancilla. It can be implemented with a non-Gaussian gate~\cite{LPSW17,MPSW15}, if $|0\rangle, |1\rangle$ represent logical qubits realized by a pair of qumodes $|01\rangle = \hat{b}^\dagger |00\rangle, |10\rangle = \hat{a}^\dagger |00\rangle$, so that $Z_A = \hat{b}^\dagger \hat{b} - \hat{a}^\dagger \hat{a}$. We obtain
\be \frac{\mathbb{I} + Q}{2}e^{i\frac{\pi\delta}{2} p_R \tilde{p}_R \hat{N}}|\psi\rangle|\phi\rangle |0\rangle_A + \frac{\mathbb{I}- Q}{2} e^{-i\frac{\pi\delta}{2} p_R \tilde{p}_R \hat{N}}|\psi\rangle|\phi\rangle |1\rangle_A ~. \ee
Finally, we perform a projective measurement on the ancilla projecting it onto $|+\rangle_A$, resulting in the desired state $Q^{\delta p_R \tilde{p}_R\hat{N}} |\psi\rangle|\phi\rangle$. This projection is successful 50\% of the time, as is easily verified.

We can now implement
\be\label{eqA9} e^{-i\gamma H p_R\tilde{p}_R\hat{N}} \approx e^{-i\gamma \epsilon \tilde{H} p_R\tilde{p}_R\hat{N}} \approx \left( \prod_j e^{-i\frac{\gamma \epsilon}{M} {H}_j p_R\tilde{p}_R\hat{N}} \right)^M\ee
using $Q^{\delta p_R \tilde{p}_R\hat{N}}$, where $\delta = \frac{2\gamma\epsilon}{\pi M}$. Let $|\psi\rangle = |1\rangle |\chi\rangle$, initially. Let $P$ be any permutation matrix, so that by repeatedly acting with $P$ on $|1\rangle$, we span all states $|j\rangle$, $j=1,\dots, j_{\text{max}}$. Then
\be\label{eqA9a} \left( Q^{\delta p_R \tilde{p}_R\hat{N}} \ P\otimes\mathbb{I}\right)^{j_{\text{max}}} |\psi\rangle = \mathbb{I}\otimes \prod_j e^{-i\frac{\gamma \epsilon}{M} {H}_j p_R\tilde{p}_R\hat{N}} |\psi\rangle~.\ee
Having constructed the unitary \eqref{eqA9a}, we may implement the unitary \eqref{eqA9} by repeating the above process $M$ times. We will use this construction
to implement the unitary \eqref{equ13} 
following~\cite{RSL16}. Let $\rho = |\chi\rangle\langle \chi |$ (eq.\ \eqref{eq:chi-y-k}) be the state on which \eqref{equ13} will act. We introduce the symmetric state
\be |s\rangle = \frac{1}{2\sqrt{N}} \sum_{x=0}^{2N-1} |x\rangle. \ee
We then act on the state $|s\rangle\langle s| \otimes \rho$ with the unitary \eqref{eqA9a}, and trace over the degrees of freedom of the register in which $|s\rangle$ resides.
We obtain
\be\label{eq28} \text{tr} \left[ e^{-i\frac{\gamma}{M} H p_R\tilde{p}_R\hat{N}} \ |s\rangle\langle s| \otimes \rho\ e^{i\frac{\gamma}{M} H p_R\tilde{p}_R\hat{N}}\right] \approx \mathcal{U}_M \ \rho\ \mathcal{U}_M^\dagger~, \ee
where $\mathcal{U}_M = e^{-i\frac{\gamma}{4MN} \hat{K}\hat{N} p_R\tilde{p}_R}$. Thus, the above procedure yielded an evolution involving the desired matrix $\hat{K}$ from the quadratically enlarged, but dynamically simpler, matrix $H$ containing all the entries of $\hat{K}$ (eq.\ \eqref{eq12}).

The error in \eqref{eq28} is $\epsilon_M \lesssim \frac{\gamma^2}{M^2\xi^4} \| \hat{K} \|^2$,
where $\| \hat{K} \|$ is the magnitude of the largest matrix element of $\hat{K}$, and we used $|\lambda| \le 4N \| \hat{K} \|$, where $\lambda$ is any eigenvalue of $\hat{K}$. 

By repeating this process $M$ times, we arrive at the desired result \eqref{equ13} (since $\mathcal{U} \approx (\mathcal{U}_M)^M$). The cumulative error is $\epsilon = M\epsilon_M \lesssim \frac{\gamma^2}{M\xi^4} \| \hat{K} \|^2$. For a large number of steps, only large enough eigenvalues contribute, specifically, $\gamma |\lambda|/N \gtrsim \xi^2$.
Let us choose a small enough squeezing parameter $\xi$ (restricted by current technology) and a large enough adjustable parameter $\gamma$, so that the error introduced by restricting to relatively large eigenvalues is $\epsilon$, i.e.,
\be\label{eq29} \gamma \sim \frac{\xi^2}{\epsilon} \ee
It should be noted that the smallest eigenvalue of $\hat{K}$ can also be controlled to a certain extent by increasing the variance noise $\sigma^2 \mathbb{I}$, so that $\| \hat{K} \| \gtrsim \sigma^2$, which may relax the constraint \eqref{eq29} on $\gamma$.

It follows that the number of oracle calls required for the algorithm is $M \lesssim \frac{\gamma^2}{\epsilon\xi^4} \| \hat{K} \|^2 \sim \| \hat{K} \|^2 / \epsilon^3$.
If $K$ is a low-rank matrix which is dense with relatively small matrix elements, then $\| \hat{K} \| \sim \mathcal{O} (\text{poly\ log} N)$ \cite{RSL16}. Moreover, if we are interested in errors $1/\epsilon \sim \mathcal{O} (\text{poly\ log} N)$, then also the number of oracle calls $M \sim \mathcal{O} (\text{poly\ log} N)$.

The above considerations constrain the parameters $\xi$ and $\gamma$ to be in a range that facilitates an accurate calculation of the mean. Indeed, we obtain
\be \mathcal{U} |\chi\rangle = \frac{1}{\sqrt{2}} \left( |\bm{y}\rangle |0\rangle + e^{i\gamma \frac{\hat{K}}{4N} p_R \tilde{p}_R} |\bm{k}_\ast\rangle|1\rangle \right) |\Phi_{R}(\xi)\rangle~.  \ee
Next, we measure the $q$ quadrature, $q_R$ and $\tilde{q}_R$, respectively, of the resource modes. If the outcome is $(q_R, \tilde{q}_R)$ with $|q_R| , |\tilde{q}_R| \lesssim \xi$, then the state is projected onto
$\Pi_\xi  \mathcal{U}|\chi\rangle$, where
\be \Pi_\xi = \int_{-\xi}^{\xi} dq_R\int_{-\xi}^{\xi} d\tilde{q}_R |q_R, \tilde{q}_R\rangle\langle q_R,\tilde{q}_R|~.  \ee
As shown below, this results in a state which is independent of the resource measurement outcomes to a good approximation. Therefore, the probability that the resource measurement successfully implements $\Pi_\xi$ is
\be \left( \frac{1}{\sqrt{\pi}\xi}\int_{-\xi}^\xi dq e^{-q^2/\xi^2} \right)^2 =  \text{erf}^2 (1) \ee
or numerically 71\%.

After a straightforward calculation, we obtain
\be \langle q_R,\tilde{q}_R| e^{i\gamma \frac{\hat{K}}{4N} \hat{N} p_R \tilde{p}_R} |\Phi_{R}(\xi)\rangle
\propto \frac{ e^{ - \frac{\xi^2 (q_R^2+\tilde{q}_R^2)+2i\gamma q_R\tilde{q}_R\frac{\hat{K}}{4N}\hat{N}}{2(\xi^4+\gamma^2 \frac{\hat{K}^2}{16N^2} \hat{N}^2)}} }{\sqrt{\xi^4+\gamma^2 \frac{\hat{K}^2}{16N^2} \hat{N}^2}}~.
\ee
where the remaining operators $\hat{K}$ and $\hat{N}$ act on the Hilbert space of the state of our system \eqref{eq:yk}, with $\hat{N}$ acting on its last qubit and $\hat{K}$ on the rest of the qubits.
Choosing $\xi$ to be small enough and $\gamma$ to be large enough, as outlined above, we may approximate 
\be  \frac{\langle q_R,\tilde{q}_R| e^{i\gamma \hat{K}  p_R \tilde{p}_R} |\Phi_{R}(\xi)\rangle}{\langle q_R,\tilde{q}_R |\Phi_{R}(\xi)\rangle} \approx \frac{\xi^2}{\gamma}  \left( \frac{\hat{K}}{4N} \right)^{-1}~.\ee
The resource modes decouple, and the remaining projected state $|\hat{\chi}(\bm{y},\bm{k}_\ast,s)\rangle$ is approximately
\be  |\hat{\chi}\rangle \approx |\bm{y}\rangle |0\rangle + \frac{\xi^2}{\gamma} \hat{K}^{-1} |\bm{k}_\ast\rangle|1\rangle ~.  \ee
Next, we measure $Z$ on the first qubit of the $n+1$ qubit system and $X$ on the appended (last) qubit. The expectation value of $\frac{\mathbb{I}+Z}{2}\otimes X $ for the state $|\hat{\chi}\rangle$ is
\begin{align} \left\langle \hat{\chi}\left| \frac{\mathbb{I}+Z}{2}\otimes X\right|\hat{\chi} \right\rangle &= \frac{2\xi^2}{Nc(y)c(k_\ast) \gamma}\bm{y}^T K^{-1} \bm{k}_\ast\nonumber\\ &= \frac{2\xi^2\bm{y} \cdot \tilde{\bm{k}}_\ast}{Nc(y)c(k_\ast) \gamma},  
\end{align}
from which we easily deduce the mean $\bm{y} \cdot \tilde{\bm{k}}_\ast$:
\be 
\bm{y} \cdot \tilde{\bm{k}}_\ast=\frac{N\gamma}{2\xi^2}c(y)c(k_\ast) \left\langle \hat{\chi}\left| \frac{\mathbb{I}+Z}{2}\otimes X\right|\hat{\chi} \right\rangle. \ee

\subsection{Efficient computation of variance}\label{sec:algo-cov}
To calculate the variance, we need to be able to efficiently compute $\bm{k}_\ast\cdot\tilde{\bm{k}}_\ast = \bm{k}^T K^{-1} \bm{k}_\ast$ \eqref{eq:variance_formula} as $k_{\ast \ast}$ is given. The calculation of $\bm{k}_\ast\cdot\tilde{\bm{k}}_\ast$ follows the same lines of the calculation of the mean with $\bm{y}$ replaced by $\bm{k}_\ast$ in \eqref{eq:yk} and \eqref{eq:chi-y-k}. Because of this replacement, we need to consider the $(n+1)$-qubit state $|\bm{k}_\ast\rangle$ corresponding to a $2N$-dimensional vector.

To this system, we append a qubit in the state $\frac{1}{\sqrt{2}} (|0\rangle + |1\rangle)$ as well as two resource modes in the squeezed state
\eqref{eq18},
thus forming the state
\be |\chi(\bm{k}_\ast,\bm{k}_\ast,\Phi(\xi))\rangle = \frac{1}{\sqrt{2}}|\bm{k}_\ast\rangle (|0\rangle + |1\rangle) |\Phi_{R} (\xi)\rangle~. \ee
Following the same steps as that for the calculation of the mean, we 
 obtain
\begin{align} \left\langle \hat{\chi}\left| \frac{\mathbb{I}+Z}{2}\otimes X\right|\hat{\chi} \right\rangle &= \frac{2\xi^2}{N\gamma c^2(k_\ast) }\bm{k}_\ast^T K^{-1} \bm{k}_\ast\nonumber\\ & = \frac{2\xi^2\bm{k}_\ast \cdot \tilde{\bm{k}}_\ast}{N\gamma c^2(k_\ast) }~,  
\end{align}
from which we easily deduce the variance $\tilde{k}_{\ast\ast}$:
\be 
\tilde{k}_{\ast\ast}={k}_{\ast\ast} -\frac{N\gamma}{2\xi^2}c^2(k_\ast) \left\langle \hat{\chi}\left| \frac{\mathbb{I}+Z}{2}\otimes X\right|\hat{\chi} \right\rangle~. \ee
Given that both $\bm{y}$ and $\bm{k}_\ast$ are sparse, and $K$ is well-conditioned, we have an efficient way of computing both the mean and variance function of Gaussian process regression. In this situation, because we are making use of a quantum linear systems algorithm, we achieve an exponential speed-up over its classical counterpart.

\section{Conclusion}\label{sec:con}

We presented a continuous-variable quantum system assisted Gaussian process regression algorithm that offers the potential of an exponential speed-up over classical techniques. It generalized the result given in \cite{ZFF15} where the authors had initially considered the application of quantum systems of linear equations algorithm \cite{HHL09} to Gaussian process regression using discrete-variable quantum systems. The application of the such HHL algorithm constraints the matrix $K$~\eqref{eq:matrix-K} related to Gaussian processes to be well-conditioned. $K$ needs to be robustly invertible \cite{Aaronson2015}, which restricts the condition number $\kappa$ to remain low even as $N$ increases. We can make $K$ robust by increasing the variance noise ($\sigma^2\mathbb{I}$) so that $\lambda_{\text{min}}$ remains above a certain threshold ($\lambda_{\text{min}} \gtrsim \sigma^2$). This dilution trick would work only if the statistical properties of the concerned model are not significantly altered. 

In \cite{RSL16}, the authors provided a method for non-sparse matrices in an oracular setting which required only one-sparse simulation techniques. We made use of this method to encode $K$ for the computation of the mean and covariance function of a Gaussian process regression model. Our presented method provides a continuous-variable quantum-assisted singular value decomposition of non-sparse low rank matrices. This hints at applications of our technique to subroutines beyond quantum systems of linear equations algorithms. 

\bigskip
\acknowledgements
We thank Patrick Rebentrost for helpful comments and feedback. SD acknowledges support from
the LSU Graduate School Economic Development Assistantship. 
GS acknowledges
support from the U.S.\ Office of Naval Research
under award number N00014-15-1-2646.

\bibliographystyle{unsrt}
\bibliography{gml}

\end{document}